\documentclass[12pt]{article}

\input epsf

\def\fnote#1#2{\begingroup\def\thefootnote{#1}\footnote{#2}
    \addtocounter{footnote}{-1}\endgroup}

\def\Email{makoto.natsuume@kek.jp}

\newcommand{\be}{\begin{equation}}
\newcommand{\ee}{\end{equation}}
\newcommand{\bea}{\begin{eqnarray}}
\newcommand{\eea}{\end{eqnarray}}

\newcommand{\x}{\times}

\newcommand{\eq}[1]{(\ref{eq:#1})}
\newcommand{\sect}[1]{Sec.~\ref{sec:#1}}
\newcommand{\app}[1]{App.~\ref{sec:#1}}
\newcommand{\fig}[1]{Fig.~\ref{fig:#1}}

\def\IR{\relax{\rm I\kern-.18em R}}

\newcommand{\ap}{$\alpha'$}

\newcommand{\enhancon}{enhan\c{c}on }
\newcommand{\del}{\partial}
\newcommand{\dX}{\dot{X}}
\newcommand{\hmu}{\hat{\mu}}
\newcommand{\hnu}{\hat{\nu}}

\newcommand{\ls}{l_s}
\newcommand{\gA}{g_A}
\newcommand{\gh}{g_h}
\newcommand{\vA}{v_A}
\newcommand{\VA}{V_A}
\newcommand{\vh}{v_h}
\newcommand{\Vh}{V_h}
\newcommand{\dis}[1]{\displaystyle{#1}}

\newcommand{\lt}{\hat{0}}
\newcommand{\lz}{\hat{4}}
\newcommand{\lr}{\hat{r}}
\newcommand{\mt}{\hat{\tau}}
\newcommand{\mz}{\hat{x}}
\newcommand{\mr}{\hat{\rho}}

\begin{document}

\begin{flushright}
%        \today\\
%        {\it draft version}\\
       KEK-TH-789\\
        hep-th/0111044\\
\end{flushright}

\vspace{18pt}

\begin{center}
{\large \bf The Heterotic Enhan\c{c}on} 

\vspace{16pt}
Makoto Natsuume \fnote{*}{\Email}

\vspace{16pt}
{\sl Theory Division\\
Institute of Particle and Nuclear Studies\\
KEK, High Energy Accelerator Research Organization\\
Tsukuba, Ibaraki, 305-0801 Japan}

\vspace{12pt}
{\bf ABSTRACT}

\vspace{12pt}
\begin{minipage}{4.8in}
The \enhancon mechanism is studied in the heterotic string theory. We consider the $N_L=0$ winding strings with momentum (NS1-W*) and the Kaluza-Klein dyons (KK5-NS5*). The NS1-W* and KK5-NS5* systems are dualized to the D4-D0* and D6-D2* systems, respectively, under the $d=6$ heterotic/IIA S-duality. The heterotic form has a number of advantages over the type IIA form. We study these backgrounds and obtain the \enhancon radii by brane probe analysis. The results are consistent with S-duality.
\end{minipage}

\end{center}

%\begin{flushright}
%       PACS codes: 11.25.-w, 04.20.Dw, 04.50.+h, 04.65.+e\\
%\end{flushright}

\vfill
\pagebreak

\baselineskip=16pt
%\parskip=12pt

%%%%%%%%%
\section{Introduction}\label{sec:intro}
%%%%%%%%%

Recently, there have been many new examples to resolve singularities in string theory \cite{Johnson:2000qt}-\cite{Adams:2001sv}. (See Ref.~\cite{Natsuume:2001ba} for a review of the singularity problem in string theory.) They typically resolve naked singularities.

These works originated from the study of gauge/gravity duals. As a result, many of them are systems with D3-branes, and they have less supersymmetry than the maximal one. 
The original Maldacena duality is the duality between the type IIB string on $AdS_5 \times S^5$ and $D=4$, ${\cal N}=4$ supersymmetric Yang-Mills theory. 
%(See Ref.~\cite{Aharony:2000ti} for a review about the duality.)  
It is of course very important to study other duals with less supersymmetry in order to study a more realistic gauge theory and its physics.
%({\it e.g.}, confinement). 
This is obviously a topic many people have considered, and thus we have many examples. In this case, the background is often $AdS_5 \times X$ asymptotically, where $X$ is a more general transverse space than the original $S^5$. So, naive deformations of $AdS_5$ tend to produce singularities, but stringy physics cures the problem. 

The singularity in the original geometry is often naked. Moreover, they are ``unphysical." It seems unlikely that one can interpret them as physically reasonable sources. This contrasts with traditional approaches of singularity resolution ({\it e.g.}, \ap-corrections and orbifolds), where one interprets the singularities as sources. A typical unphysical behavior is that gravitational force becomes repulsive near the singularity. This type of singularity is known as ``repulson." %(The Reissner-Nordstr\"{o}m solution has the same behavior inside the inner horizon.) 
They are unphysical just like the $M<0$ Schwarzschild \cite{Horowitz:1995ta}, but the final geometries usually do not have the problem ({\it e.g.}, all enhan\c{c}on examples known to date, the Klebanov-Strassler solution, and some of ``transgression" examples).
%In the terminology of \sect{horowitz-myers}, they ``excise" singularities, not ``resolve" singularities. 

The \enhancon mechanism \cite{Johnson:2000qt} is one of the most studied mechanisms and is applied to many systems \cite{Buchel:2001cn}-\cite{Merlatti:2001gd}. The prototypical example, the K3-wrapped D$p$-branes for $p=4,\ldots,7$, has been studied further. (See Refs.~\cite{Johnson:2001wm}-\cite{Wijnholt:2001us} for recent discussion.)

For example, the K3-wrapped D6-brane induces a negative D2 tension and charge \cite{Bershadsky:1996qy}-\cite{Bachas:1999um}. (For that reason, the system is sometimes denoted as D6-D2*.) The naive supergravity solution has a repulson singularity. However, the constituents, the wrapped D6-branes, become massless before reaching the singularity. Thus, the naive supergravity analysis breaks down, and one must take into account stringy physics. Their conclusion is that constituent branes do not form the singularity. Rather, they stay at a finite radius $r=r_e$ and form a ``shell." Since all the sources are at this radius, the geometry is flat in the interior. The claim is based on the fact that there is no consistent way to move the branes to $r<r_e$ further without contradicting with supersymmetry. This claim is further supported by the ${\cal N}=2$ supersymmetric gauge theory analysis. In the $d=6$ language, a D4-D0* system, which is also massless at $r=r_e$, is a W-boson, and the D6-D2* system corresponds to a SU(2) monopole. The fact that the D6-D2* system cannot move inside the \enhancon radius has the gauge theory interpretation that the size of the monopole is inversely proportional to the W-boson mass; hence, the brane cannot be localized as $r \rightarrow r_e$.

Now, a different string theory realizes this phenomenon by a different mechanism. The type IIA string on K3 is S-dualized to the heterotic string on $T^4$, and the phenomenon (for $p=4$) is a consequence of the self-dual radius in this heterotic form; the BPS winding string with $N_L=0$ becomes massless when the $S^1$ radius the string wraps becomes the string scale $\ls$.%
\footnote{The terms ``BPS string with $N_L=0$" and ``BPS $(n,w)=(1,1)$ string" are used interchangeably. Since we consider only the BPS systems, the term BPS is often omitted for brevity. In the heterotic form, this phenomenon should not be called as an enhan\c{c}on, since these theories generally have no overlapping region of validity (this is because S-duality is involved). However, we will continue to refer to it as an \enhancon mechanism for simplicity.}
In fact, the repulson singularity is first discussed in this context \cite{Behrndt:1995tr}-\cite{Cvetic:1995mx}. As another example, a different dual system consists of D$(p-3)$-branes ending on a pair of NS5-branes. 
In this system, the phenomenon is realized by Myers' effect, and it has been studied in detail \cite{Johnson:2001bm}.  

We study the \enhancon mechanism in the heterotic form in detail. The heterotic form has some advantages over the type IIA form. The massless BPS winding string is just a familiar perturbative effect. The heterotic form is also somewhat cleaner since \ap-corrections are under better control.
In the type IIA form, the naive D4-D0* solution is just a leading order solution in \ap, so it is not clear whether the repulson geometry is really singular. (However, if \ap-corrections resolved the singularity, one would be left with the unphysical region near the singularity.) In contrast, the solution for the $(1,1)$ string is known to be exact to all orders in \ap \cite{Horowitz:1994ei}-\cite{Horowitz:1995rf}, so the singularity survives the \ap-corrections (apart from the \ap-corrections due to source terms \cite{Tseytlin:1995uq}). Also, in the type IIA string, the curvature is actually strong near the \enhancon radius, which complicates the situation (\sect{geometry}), whereas the curvature remains weak near the \enhancon radius in the heterotic form. Finally, in order to obtain the probe action for $r<r_e$, Ref.~\cite{Johnson:2000qt} takes the absolute value of the probe Dirac-Born-Infeld action by hand; in the heterotic form, the absolute value appears naturally (\sect{NS1Wprobe}). In short, the heterotic form strongly supports the picture of Ref.~\cite{Johnson:2000qt}.
This work also provides an interesting application of the Kaluza-Klein monopole action \cite{Bergshoeff:1997gy}-\cite{Eyras:1998hn}, which is rarely discussed in the literature. For other works on the heterotic form, see Refs.~\cite{Wijnholt:2001us,Krogh:1999qr}.

The heterotic form has some disadvantages as well; the analysis in the heterotic string is more involved. The probe analysis is not so simple as the D-brane probe analysis. The D2*-brane effect comes from the higher order corrections in the D-brane action, but the similar corrections for the Kaluza-Klein monopole action are less known. The $d=10$ singularity of the $(1,1)$ string is not a standard curvature singularity, but a parallelly propagated curvature singularity (\sect{repulson} and \app{appA}).

The organization of the present paper is as follows. In the next section, we describe the backgrounds we study. We show that they are S-dual to the type IIA backgrounds and that they have repulson singularities even from the $d=10$ point of view. We also give the interpretation of those backgrounds \cite{Behrndt:1995tr}-\cite{Cvetic:1995mx}, \cite{Sen:1997zb}. In \sect{probes}, we carry out the probe analysis in those backgrounds. It is also shown that both the ``W-boson" and ``SU(2) monopole" become massless simultaneously in those backgrounds, which is crucial for the validity of the \enhancon mechanism. As in the type IIA cases, the \enhancon radii can actually be identified from physical reasoning only. Thus, we first present the heuristic analysis in \sect{heuristic} to identify the \enhancon radii, and then we justify the results using the probe actions. This is useful since the probe analysis is somewhat more involved than the type IIA cases.

%%%%%%%%%
\section{Heterotic Repulsons}\label{sec:backgrounds}
%%%%%%%%%

%%------------------------------------------------
\subsection{Geometry and Interpretation}\label{sec:geometry}
%%------------------------------------------------

We consider a system of $N$ winding strings with momentum, and a system of $N$ Kaluza-Klein dyons \cite{Sen:1997zb}. We denote our backgrounds as NS1-W* and KK5-NS5*, respectively, to distinguish them from the standard solutions. However, this is somewhat misleading because the NS1-W* system really corresponds to the $(n,w)=(\pm 1, \pm 1)$ string \cite{Behrndt:1995tr}-\cite{Cvetic:1995mx}. Denote a world-volume direction of a brane by $\x$ and a transverse direction by $-$. Then, the brane configurations we consider in the type IIA string are given by
\be
\begin{array}{lcc|ccccc|cccc}
			&& 0 & 1 & 2 & 3 & 4 & 5 & 6 & 7 & 8 & 9  \\ \hline
\mbox{D0*} 	&& \x& - & - & - & - & - & - & - & - & -  \\ 
\mbox{D4} 	&& \x& - & - & - & - & - & \x& \x& \x& \x \\ \hline
\mbox{D2*} 	&& \x& - & - & - & \x& \x& - & - & - & -  \\ 
\mbox{D6}	&& \x& - & - & - & \x& \x& \x& \x& \x& \x 
\end{array}
\ee
whereas the brane configurations in the heterotic string are given by
\be
\begin{array}{lcc|ccccc|cccc}
			&& 0 & 1 & 2 & 3 & 4 & 5 & 6 & 7 & 8 & 9  \\ \hline
\mbox{NS1} 	&& \x& - & - & - & - & - & - & - & - & \x \\
\mbox{W*} 	&& \x& - & - & - & - & - & - & - & - & z  \\ \hline
\mbox{KK5} 	&& \x& - & - & - & \x& \x& \x& \x& \x& z  \\
\mbox{NS5*}	&& \x& - & - & - & \x& \x& \x& \x& \x& - 
\end{array}
\label{eq:heterotic}
\ee
The $z$ is the direction of momentum, or the transverse direction of the Kaluza-Klein monopole with no Goldstone mode. The $x^6, \ldots, x^9$ directions are compact in both cases. The D$p$-D$(p-4)^*$ systems are compactified on K3, and the NS-brane systems are compactified on $T^3 \times S^1$, where $S^1$ is the $z=x^9$ direction.
There are basically three relative alignments of the NS1-W* and KK5-NS5* branes. The NS1-W* system may be aligned
\begin{enumerate}
\item with the world-volume directions of the KK5-NS5* system ($x^4, \ldots, x^8$), 
\item with the transverse directions ($x^1, \ldots, x^3$), or 
\item with the isometry direction $z$. 
\end{enumerate}
The important configuration is the last of these since both the NS1-W* and KK5-NS5* branes give extra massless spectra from the configuration.

The D$p$-D$(p-4)^*$ solution is given by
\bea
ds^2 &=& 
Z_{p-4}^{-1/2}Z_{p}^{-1/2} \eta_{mn} \, dx^m dx^n
	+ Z_{p-4}^{1/2}Z_{p}^{1/2} dx^i dx^i 
	+ \vA^{1/2} Z_{p-4}^{1/2}Z_{p}^{-1/2} ds^2_{\rm K3},
\nonumber \\
e^{-2\phi} &=& \frac{1}{\gA^2} Z_{p-4}^{-(7-p)/2} Z_{p}^{(p-3)/2}, \nonumber \\
C_{p-3} &=& \frac{1}{\gA} (Z_{p-4}^{-1}-1) \,
dx^0 \wedge dx^{10-p} \wedge \cdots \wedge dx^5, \nonumber \\
C_{p+1} &=& \frac{1}{\gA} (Z_{p}^{-1}-1) \,
dx^0 \wedge dx^{10-p} \wedge \cdots \wedge dx^9.
\label{eq:DpD(p-4)}
\eea
Here, $m, n$ run over the directions tangent to all the branes, and $i$ runs over the directions transverse to all branes. The metric $ds^2_{\rm K3}$ is a K3 surface metric of volume $(2\pi \ls)^4$, and the K3 volume is
\bea
\VA &=& (2\pi\ls)^4 \vA, \nonumber \\
\vA &=& R_{6A} \ldots R_{9A}
\eea
in the orbifold limit. Namely, we measure the compact volume in units of $2\pi \ls$, which makes T-duality rules simple. All compact spatial coordinates run from 0 to $2\pi\ls$. (All world-volume coordinates also run from 0 to $2\pi\ls$.) The coupling constant in the type IIA (heterotic) string is denoted as $g_A$ ($g_h$). Then, the harmonic functions are given by
\renewcommand{\arraystretch}{2.2}
\be
\begin{array}{cclccl}
Z_0 &=& \dis{1-\frac{|r_0|^3}{r^3}},& 
r_0^3 &=& \dis{\frac{\pi \gA N \ls^3}{\vA}}, \\
Z_4 &=& \dis{1+\frac{r_4^3}{r^3}},&
r_4^3 &=& \pi \gA N \ls^3
\end{array}
\ee
\renewcommand{\arraystretch}{1}%
for the D4-D0* case ($r^2=x^ix^i$), and 
\renewcommand{\arraystretch}{2}
\be
\begin{array}{cclccl}
Z_2 &=& \dis{1-\frac{|r_2|}{r}}, & 
r_2 &=& \dis{\frac{\gA N \ls}{2\vA}}, \\
Z_6 &=& \dis{1+\frac{r_6}{r}}, &
r_6 &=& \dis{\frac{\gA N \ls}{2}}
\end{array}
\ee
\renewcommand{\arraystretch}{1}%
for the D6-D2* case. 

Similar to Ref.~\cite{Johnson:2000qt}, we simply analytically continue a harmonic function of a brane. The heterotic/IIA S-duality guarantees that such systems should exist, but the interpretation is given later. The branes with * are chosen so that the total mass is positive when the $S^1$ radius is large. Then, the NS1-W* solution is given by
\bea
ds^2 &=& 
f_1^{-1} \{ -dt^2+R^2dz^2+(f_n-1)(dt-Rdz)^2 \}
	+ dx^i dx^i 
	+ v_3^{2/3} ds^2_{T^3},
\nonumber \\
e^{-2\phi} &=& \frac{1}{\gh^2} \, f_1, \nonumber \\
B_{2} &=& (f_1^{-1}-1) \, dt \wedge (Rdz),
\label{eq:NS1W}
\eea
where
\renewcommand{\arraystretch}{2.2}
\be
\begin{array}{cclccl}
f_1 &=& \dis{1+\frac{r_1^3}{r^3}}, & 
r_1^3 &=& \dis{\frac{\pi \gh^2 N \ls^3}{\vh} R}, \\
f_n &=& \dis{1-\frac{|r_n|^3}{r^3}}, & 
r_n^3 &=& \dis{\frac{\pi \gh^2 N \ls^3}{\vh} \frac{1}{R}}.
\end{array}
\ee
\renewcommand{\arraystretch}{1}%
The torus volume is defined similar to the type IIA cases; $ds^2_{T^3}$ is the  $T^3$-metric of volume $(2\pi\ls)^3$, $R = R_{9h}$, and 
\bea
\Vh &=& (2\pi\ls)^4 \vh, \nonumber \\
\vh &=& Rv_3.
\eea

The KK5-NS5* solution is given by
\bea
ds^2 &=& 
\eta_{mn} dx^m dx^n
	+ f_5 \left\{ f_K (dr^2+r^2d\Omega_2^2) 
	+ f_K^{-1} R^2 (dz + A_\varphi d\varphi)^2 \right\}
	+ v_3^{2/3} ds^2_{T^3},
\nonumber \\
e^{-2\phi} &=& \frac{1}{\gh^2} \, f_5^{-1}, \nonumber \\
A_1 &=& \frac{r_K}{R} (1 - \cos\theta) \, d\varphi, \nonumber \\
B_2 &=& - r_5 (1 - \cos\theta) \, d\varphi \wedge (Rdz), 
\label{eq:KK5NS5}
\eea
where
\renewcommand{\arraystretch}{2}
\be
\begin{array}{cclccl}
f_5 &=& \dis{1-\frac{|r_5|}{r}}, &
r_5 = \dis{\frac{N \ls}{2} \frac{1}{R}}, \\
f_K &=& \dis{1+\frac{r_K}{r}}, & 
r_K = \dis{\frac{N \ls}{2} R}.
\end{array}
\ee
\renewcommand{\arraystretch}{1}%

The gravitational wave W* is no more an anti-wave than the D2*-brane is an anti-D2-brane. The anti-wave is just a parity reflection $z \rightarrow -z$ of the original system. The solution does not change and is uninteresting. Rather, the NS1-W* system is interpreted as a true $(n,w)=(1,1)$ string \cite{Behrndt:1995tr}-\cite{Cvetic:1995mx}. In \sect{NS1Wprobe}, we indeed see that a $(1, 1)$ probe string has a flat potential in this background, but a $(-1, 1)$ probe string does not. Also, recall the mass-shell conditions for the heterotic string:
\bea
\mbox{right-moving (NS-sector): } \quad m^2 &=& P_R^2 + \frac{4}{\ls^2}(N_R-\frac{1}{2}), \nonumber \\
\mbox{left-moving: } \quad m^2 &=& P_L^2 + \frac{4}{\ls^2}(N_L-1), 
\label{eq:mass_shell}
\eea
where $P_{L,R} = (n/R \pm wR)/\ls$. The BPS states are those with one world-sheet fermion $\psi_{-1/2}^\mu$ excited or $N_R=1/2$. Thus, the $(1,1)$ string has the mass $m^2 = (R-1/R)^2/\ls^2$. This mass formula coincides with the NS1-W* mass $M \sim r_1^3 - r_n^3 \sim R-1/R$. Thus, roughly speaking, the effective negative energy from W* comes from the Casimir energy. At the self-dual radius $R=1$, the massless spectrum is given by $(N_L,n,w) = (1, 0, 0), (0, \pm1, \pm1)$, and they form an enhanced SU(2) gauge symmetry. 

The NS5*-brane tension and charge are induced by the KK5-brane curvature due to the heterotic anomaly equation \cite{Krogh:1999qr,Sen:1997zb}:
\be
H_3 = dB_2 + \frac{\ls^2}{4} \left( \omega_3^L(\Omega) - \frac{1}{30} \omega_3^{YM}(A) \right),
\label{eq:anomalyeq}
\ee
where $\omega_3^L$ and $\omega_3^{YM}$ are the Lorentz and Yang-Mills Chern-Simons three-forms. The Lorentz Chern-Simons term induces $(-1)$ unit of NS5-brane charge since a Kaluza-Klein monopole has $(+1)$ unit of gravitational instanton number. The backreaction of the NS5*-geometry on Eq.~\eq{anomalyeq} is subleading in $1/R$ (higher orders in \ap). Since the system is a BPS object, its tension should also be modified appropriately.

The above result, however, assumes weak curvature. Since the heterotic anomaly equation \eq{anomalyeq} could get higher order corrections in \ap, the issue of \ap-corrections may arise. If \ap-corrections become important, the background solution \eq{KK5NS5} could not be reliable as well. However, the curvature remains weak in the region we are interested in, namely near the \enhancon radii (\sect{heuristic}). More importantly, the weak curvature result continues to hold due to the standard charge quantization and the BPS condition \cite{Krogh:1999qr}. 

A similar point arises for type IIA cases as well. Near the \enhancon radii, the size of K3 is about the string scale $\ls$. Since K3 is curved, the curvature near the \enhancon radius generally becomes strong, so the \ap-corrections are not negligible. This could have changed the results of Ref.~\cite{Johnson:2000qt}. The K3-wrapped D6-brane has $(-1)$ unit of D2-brane charge, but the analysis takes into account only the leading \ap-corrections in the Dirac action and in the Chern-Simons coupling. The actions should have higher order corrections in \ap, and those corrections may invalidate the weak curvature result. In particular, the naive D6-D2* probe action would not make sense. However, the K3-wrapped D6-brane continues to have $(-1)$ unit of D2-brane charge even in the large curvature region due to the charge quantization. Thus, the effective probe action itself is still reliable, but the D6-D2* background may be modified.

%%------------------------------------------------
\subsection{Repulson Singularities}\label{sec:repulson}
%%------------------------------------------------

The heterotic systems are S-dual to the type IIA ones, so their $d=6$ geometries have the same repulson behaviors (in the Einstein metric). Here, we examine repulson singularities from the $d=10$ point of view. It is sometimes stated that the NS1-W* geometry is regular at $r=r_n$, but we will see this is not the case.

For the KK5-NS5* background, the Ricci scalar is
\be
R_{\rm KK5-NS5^*} = \frac{3r_5^2}{2(r-r_5)^3(r+r_K)},
\ee
so it is singular at $r=r_5$, but the Ricci scalar of the NS1-W* background is finite at $r=r_n$:
\be
R_{\rm NS1-W^*} = -\frac{63r_1^6}{2r^2(r^3+r_1^3)}.
\ee
However, one cannot immediately conclude that the NS1-W* geometry is regular at $r=r_n$. First of all, singularities at which curvature invariants diverge are not the only singularities that could appear. In general relativity, one can classify singularities as follows \cite{tipler80}:
\begin{enumerate}
\item[(i)] s.p. (scalar polynomial) curvature singularities; 
\item[(ii)] p.p. (parallelly propagated) curvature singularities;  
\item[(iii)] others (the conical singularity, Taub-NUT space, \ldots). 
\end{enumerate}
The singularity such as the KK5-NS5* system corresponds to type (i), but it is well known that the singularity of the plane wave corresponds to type (ii). All curvature invariants remain finite for the background, but it still has a diverging tidal force (see, {\it e.g.}, Ref.~\cite{Horowitz:1990bv}). This holds for more general cases, and the singularity due to momentum cannot be seen from curvature invariants only \cite{Kaloper:1997hr}. Thus, one had better be careful to check singularities when one adds momentum.  A proper way to see the singularity is to compute the Riemann tensor components in parallelly propagated frames. Note that the tidal force is proportional to the Riemann tensor from the equation of geodesic deviation. Since its components are frame-dependent, it is important to determine a natural frame to see them. One natural frame is the rest frame of an infalling observer, which is known as a parallelly propagated frame (a frame obtained by the parallel-transport along the observer's geodesic). In \app{appA}, we show that the NS1-W* geometry has a p.p. curvature singularity at $r=r_n$.

These singularities are repulson singularities from the motion of minimally-coupled massive particles. The potential wall becomes infinitely repulsive when harmonic functions of branes with * vanish. 
The existence of Killing vectors $\xi=\del_t, \del_z$ implies conserved charges $p \cdot \xi$ or $E=-p_0/m$ and $P=p_z/m$, where $E$ and $P$ are the energy and momentum along $S^1$ per unit mass, respectively. For simplicity, we consider the motion in the $z-r$ plane only and set the other conserved charges to zero. Then, the radial motion is given by
\bea
\frac{dr}{d\tau} &=& \pm \sqrt{-G^{00}G_{rr}^{-1}} \sqrt{2(\tilde{E}-V_{\rm eff})},
\nonumber \\
2V_{\rm eff} &=& (-G^{00})^{-1} (1+P^2 G^{zz}),
\label{eq:potential}
\eea
where $\tilde{E}=E^2/2$. For the NS1-W* geometry, the potential $V_{\rm eff}$ is attractive asymptotically. It begins to be repulsive at 
\be
r^3 = \frac{2R^2}{R^2-1} |r_n|^3,
\label{eq:turningpt}
\ee
and becomes infinitely strong at the singularity. The effective potential for the KK5-NS5* geometry is flat at zero momentum since $G_{00}$ is flat, but it has an infinitely repulsive potential wall at $r=r_5$ for $P \neq 0$. 

In many recent examples of the singularity resolution, the singularity in the original geometry is not only naked, but also repulsive. In fact, a repulsive singularity is timelike for a general static spherically symmetric metric (\app{appB}). Thus, it is at least ``locally naked" (like the Reissner-Nordstr\"{o}m singularity) or ``globally naked."

%%------------------------------------------------
\subsection{HET/IIA Duality}\label{sec:duality}
%%------------------------------------------------

Under the $d=6$ heterotic/IIA S-duality, the NS1-W* system is dual to the D4-D0* system, and the KK5-NS5* system to the D6-D2* system. The Kaluza-Klein procedure is standard:
\bea
\mbox{$d$-dimensional dilaton:}\quad 
\phi &\rightarrow& \sigma = \phi -\frac{1}{4} \ln \det G_c ,
\nonumber \\
\mbox{$d$-dimensional metric:}\quad
G_{\mu\nu} &\rightarrow& G_{\mu\nu}'= G_{\mu\nu} - G_{zz}^{-1}G_{\mu z}G_{\nu z}, 
\nonumber \\
\mbox{Einstein metric:}\quad
G_{\mu\nu}^E &=& e^{-\frac{4\sigma}{d-2}} G_{\mu\nu}',
\eea
where $G_c$ is the metric of a compact space. After rescaling the coordinates $t$ and $r$, the compactified type IIA branes are
\bea
ds_{E}^2 &=& -(Z_0 Z_4)^{-3/4} dt^2 + (Z_0 Z_4)^{1/4} (dr^2+r^2d\Omega_4^2),
\nonumber \\
e^{-2\sigma_A} &=& \frac{\vA}{\gA^2} (Z_0 Z_4)^{-1/2}, 
\label{eq:6dimD4D0}
\eea
\bea
ds_{E}^2 &=& (Z_2 Z_6)^{-1/4} (-dt^2+\cdots) + (Z_2 Z_6)^{3/4} (dr^2+r^2d\Omega_2^2),
\nonumber \\
e^{-2\sigma_A} &=& \frac{\vA}{\gA^2} (Z_2 Z_6)^{1/2}, 
\label{eq:6dimD6D2}
\eea
whereas the heterotic branes are
\bea
ds_{E}^2 &=& -(f_1 f_n)^{-3/4} dt^2 + (f_1 f_n)^{1/4} (dr^2+r^2d\Omega_4^2),
\nonumber \\
e^{-2\sigma_h} &=& \frac{\vh}{\gh^2} (f_1 f_n)^{1/2}, 
\label{eq:6dimNS1W}
\eea
\bea
ds_{E}^2 &=& (f_5 f_k)^{-1/4} (-dt^2+\cdots) + (f_5 f_k)^{3/4} (dr^2+r^2d\Omega_2^2),
\nonumber \\
e^{-2\sigma_h} &=& \frac{\vh}{\gh^2} (f_5 f_k)^{-1/2}.
\label{eq:6dimKK5NS5}
\eea
Becasue of the scaling, the supergravity parameters are scaled as well:
\renewcommand{\arraystretch}{2.3}
\be
\begin{array}{cclccl}%
r_0^3 &=& \pi N \ls^3 (\gA^2 \vA)^{-1/4}, &
r_4^3 &=& \dis{\pi N \ls^3 \left(\frac{\vA^3}{\gA^2}\right)^{1/4}}, \\
r_2 &=& \dis{\frac{N \ls}{2} \left(\frac{\gA^2}{\vA^3}\right)^{1/4}}, &
r_6 &=& \dis{\frac{N \ls}{2} (\gA^2 \vA)^{1/4}}, \\
r_1^3 &=& \dis{\pi N \ls^3 \left(\frac{\gh^2 R^4}{\vh}\right)^{1/4}}, &
r_n^3 &=& \dis{\pi N \ls^3 \left(\frac{\gh^2}{R^4 \vh}\right)^{1/4}}, \\
r_5 &=& \dis{\frac{N \ls}{2} \left(\frac{\vh}{\gh^2 R^4}\right)^{1/4}}, &
r_K &=& \dis{\frac{N \ls}{2} \left(\frac{R^4 \vh}{\gh^2}\right)^{1/4}}.
\end{array}
\ee
\renewcommand{\arraystretch}{1}%
The S-duality is given by $G_{\mu\nu}^E \,{\rm(IIA)} = G_{\mu\nu}^E \,{\rm (HET)}, \sigma_A = -\sigma_h$ with
\bea
\gA &=& \gh^{-1} R_{9h} v_h^{1/2}, \nonumber \\
R_{9A} &=& v_h^{-1/2} R_{9h}, \nonumber \\
R_{mA} &=& v_h^{1/2} R_{mh}^{-1}, \qquad m=6,7,8
\eea
in the orbifold limit of K3 \cite{Big}. The type IIA branes \eq{6dimD4D0} and \eq{6dimD6D2} are dual to the heterotic branes \eq{6dimNS1W} and \eq{6dimKK5NS5}, respectively, under the duality. Since the $d=6$ Einstein metric is invariant under the S-duality, both the KK5-NS5* and NS1-W* backgrounds clearly have s.p. curvature singularities at $r=r_5$ and $r=r_n$, respectively, after the reduction.

%%%%%%%%%
\section{Brane Probes}\label{sec:probes}
%%%%%%%%%

%%------------------------------------------------
\subsection{Heuristic Analysis}\label{sec:heuristic}
%%------------------------------------------------

In the D6-D2* system, the \enhancon radius can be identified from physical reasoning only. Here, we first make such a heuristic analysis for the heterotic string.

First recall the D$p$-D$(p-4)^*$ system. The tension of the $d=6$ effective $(p-4)$-brane is given by
\be
\tau = \frac{N}{\gA} (\mu_{p}\VA-\mu_{p-4}) 
= \frac{N}{\gA} \mu_{p} (\VA-V_*),
\ee
where $V_* = (2\pi)^4 \ls^4$ and 
\be
\mu_p = \frac{1}{(2\pi)^{p}\ls^{p+1}}. 
\ee
The D$p$-D$(p-4)^*$ probes all become massless at the same place $\VA=V_*$, which is crucial for the validity of the \enhancon mechanism. The $p=4,6$ systems are the W-boson and SU(2) monopole, respectively, in the $d=6$ gauge theory language. Since the monopole mass is proportional to the W-boson mass, their mass must vanish simultaneously. We see below that the NS1-W* and KK5-NS5* systems also have vanishing mass at the same radii. 

In the curved background, the K3 volume $\VA$ varies with position, so replace $\VA$ with the invariant volume $V(r)$. Then, D-brane probes become massless at $V(r)=V_*$, namely at 
\be
r = \frac{2\VA}{\VA-V_*} |r_2|,
\ee
which is the right answer from the brane probe analysis.

Now, the mass of the BPS string is given by
\be
\tau = \frac{n}{R}-wR.
\label{eq:NS1Wtension}
\ee
The KK5-NS5* tension is given by
\be
\tau = \frac{1}{\gh^2} \mu_5 (Q_K R^2-Q_5).
\label{eq:NS5KK5tension}
\ee
The mass of a KK5-brane, $m = R_4 \cdots R_8 R^2/(\gh^2 \ls)$, can be determined from the T-dual of the NS5 tension or using the Dirac quantization condition. The tensions are positive for the choice $R>1$ when $n=w=N$ and $Q_5=Q_K=N$. 

First, consider the NS1-W* background. Changing $R$ to the invariant $S^1$ radius $R(r)=R\sqrt{f_n/f_1}$, we find that the $(n,w)=(\pm1,\pm1)$ strings become massless when 
\be
f_n R^2 = f_1 
\quad {\rm or} \quad 
r^3 = \frac{2R^2}{R^2-1} |r_n|^3 = r_e^3.
\label{eq:NS1Wenhancon}
\ee
The invariant radius becomes $R(r_e) = 1$ at the \enhancon radius. Note that the KK5-NS5* tension \eq{NS5KK5tension} has the same form as the NS1-W* one \eq{NS1Wtension}, so the KK5-NS5* probe becomes massless at the same radius. Thus, the repulsive region \eq{turningpt} is actually excised by the \enhancon mechanism like the D6-D2* case. The effective potential analysis in \sect{repulson} is of course frame-dependent, and it is not very clear which frame one must choose \cite{Maldacena:2001mw,Johnson:2001wm}. If one uses the Einstein metric, not all repulsive regions are excised. (This is true for the D6-D2* case as well.)

Similarly, the invariant $S^1$ radius is $R(r)=R\sqrt{f_5/f_K}$ for the KK5-NS5* background, and the \enhancon radius is given by
\be
f_5 R^2 = f_K
\quad {\rm or} \quad 
\tilde{r}_e = \frac{2R^2}{R^2-1} |r_5|
\label{eq:KK5NS5enhancon}
\ee
for both the NS1-W* and KK5-NS5* probes.

It is easy to check that both the curvature and coupling remain weak at \enhancon radii. For the NS1-W* background,
\be
R|_{r_e} \rightarrow - \frac{63}{2 \ls^2} 
\left( \frac{R \vh}{2 \pi \gh^2 N} \right)^{2/3},
\quad 
e^{2\phi}|_{r_e} = \gh^2 \, \frac{2}{R^2+1}.
\ee
The above Ricci scalar is for the limiting case $R \gg 1$. For the KK5-NS5* background,
\be
R|_{\tilde{r}_e} 
= \frac{6}{N^2 \ls^2}\left(\frac{R^2-1}{R^2+1}\right)^4, 
\quad 
e^{2\phi}|_{\tilde{r}_e} = \gh^2 \, \frac{R^2+1}{2R^2}.
\ee
When $R>1$, they remain weak in the large-$N$ limit, $\gh \rightarrow 0, N \rightarrow \infty$ with $\gh^2 N \gg 1$. One can also check that the Riemann tensor in a p.p. frame remains weak for the NS1-W* case. On the contrary, the curvature is in general not weak in the type IIA form due to the hidden K3 curvature.

%%------------------------------------------------
\subsection{NS1-W* Probe}\label{sec:NS1Wprobe}
%%------------------------------------------------

The Lorentzian form of the $\sigma$-model action is given by
\be
S_\sigma = - \frac{1}{4\pi \ls^2} \int d^2\sigma \sqrt{-h}
\left( h^{ab}\del_a X^{\mu}\del_b X^{\nu}G_{\mu\nu} - \epsilon^{ab}\del_a X^{\mu}\del_b X^{\nu}B_{\mu\nu} \right),
\ee
where $\epsilon^{01}=1/\sqrt{-h}$ and $\epsilon_{01}=-\sqrt{-h}$. All world-volume coordinates run from 0 to $2\pi\ls$. The NS1-W probe analysis in the NS1-W background has been carried out in Refs.~\cite{Gauntlett:1994si,Callan:1996hn} in the weak field approximation. 

Our gauge choice and ansatz are as follows:
\bea
X^0 &=& \tau, \nonumber \\
X^9 &=& X^9(\tau,\sigma) = \sigma + y(\tau), \nonumber \\
%X: 0->2\pi R
%R \sigma
X^i &=& X^i(\tau), \nonumber \\
h_{\tau\tau} &=& h_{\tau\tau}(\tau), \nonumber \\
h_{\sigma\sigma} &=& h_{\sigma\sigma}(\tau), \nonumber \\
h_{\tau\sigma} &=& 0.
\eea
In this ansatz, the string has a unit winding. The conformal gauge cannot be taken because the background breaks the SO(1,1) invariance on the world-sheet due to the plane wave. (Recall $h_{ab} \propto g_{ab}$ by solving $h_{ab}$ equations of motion, where $g_{ab}$ is the induced metric.) The action then becomes
\be
S_\sigma = - \frac{2\pi\ls}{4\pi \ls^2} \int d\tau 
\left( -e^{-1} G_{\mu\nu} \dX^\mu \dX^\nu 
+ e G_{99} + 2B_{\mu9} \dX^\mu \right),
\ee
where $e \equiv \sqrt{-h_{\tau\tau}/h_{\sigma\sigma}}$. The $y$ is a cyclic coordinate, so there is a conserved momentum: 
\be
P \equiv \frac{\del L}{\del \dot{y}}
= \frac{1}{2\ls e}(2\dot{y}G_{99}+2G_{\hmu9} \dX^{\hmu} ),
\ee
where the sum on $\hmu$ does not include 9.
One can substitute back $P$ into the action and eliminate $y$ by using the Routhian \cite{mechanics}. The Routhian is defined as $ R \equiv P\dot{y}-L \rightarrow -L' $:
\bea
S'_\sigma &=& - \frac{1}{2\ls} \int d\tau 
\left\{ e^{-1} \left( -G_{\hmu\hnu} \dX^{\hmu} \dX^{\hnu}
+ \frac{(G_{\hmu9}\dX^{\hmu})^2}{G_{99}} \right) \right.
\nonumber \\
&& \left.+ e \left( G_{99}+\frac{P^2\ls^2}{G_{99}} \right) 
- \left( 2B_{\hmu9}+2\ls\frac{P G_{\hmu9}}{G_{99}} \right) \dX^{\hmu}
\right\}.
\label{eq:routhian}
\eea
As shown in Ref.~\cite{Gauntlett:1994si}, to properly take into account the Casimir energy, the correct constraint equation is as follows:
\be
T_{ab} = -\frac{1}{\ls^2}
\left(\begin{array}{cc} 
e^2 & e \\
e	& 1
\end{array}
\right).
\ee
Imposing the off-diagonal element of the constraint determines $P$:
\be
P=\frac{1}{\ls}.
\ee
Meanwhile, the diagonal element gives the correct mass-shell condition \eq{mass_shell} for the $X^{\hmu}$-independent terms in Eq.~\eq{routhian}:
\be
e \left( G_{99}+\frac{1}{G_{99}} \right) \rightarrow
e G_{99} \left( 1-\frac{1}{G_{99}} \right) ^2.
\ee
After eliminating $e$ using its equation of motion, we can expand the action in powers of transverse velocity as
\bea
\ls \times {\cal L}'_\sigma &=& 
-\left| Rf_1^{-1} - \frac{1}{R} f_n^{-1} \right|
+ \frac{1}{2R} |f_n R^2 - f_1| v^2 
\nonumber \\
&& \hspace{.5in} + R (f_1^{-1}-1) - \frac{1}{R} (f_n^{-1}-1) + \cdots
\label{eq:preNS1WinNS1W} \\
&=& -\frac{R^2-1}{R} 
+ \frac{1}{2R} (f_n R^2 - f_1) v^2 + \cdots
\label{eq:NS1WinNS1W}
\eea
in the NS1-W* background. We have taken $r>r_e$ in the last line. The potential is flat, and the effective mass takes the form as expected from the heuristic analysis \eq{NS1Wenhancon}. This analysis also shows that the NS1-W* system really corresponds to the $(n,w)=(1,1)$ string. Just as the D6-D2* analysis, the effective mass could be negative for $r<r_e$ if the second term in Eq.~\eq{preNS1WinNS1W} is not an absolute value. But it is clear in the heterotic side that this is actually spurious. The tension is always positive as expected. However, as discussed in Ref.~\cite{Johnson:2000qt}, the potential no longer cancels in this case, thus the probe at $r<r_e$ contradicts with supersymmetry. Therefore, the probe actually cannot go inside $r_e$.

Similarly, in the KK5-NS5* background (for $r>\tilde{r}_e$),
\be
\ls \times {\cal L}'_\sigma = -\frac{f_5 R^2 - f_K}{R \sqrt{f_5 f_K}} 
+ \frac{1}{2R} (f_5 R^2 - f_K) \sqrt{f_5 f_K} v^2 + \cdots.
\label{eq:NS1WinKK5NS5}
\ee
The effective mass again agrees with the heuristic analysis \eq{KK5NS5enhancon}, and the potential is of course not flat for this case.

%%------------------------------------------------
\subsection{KK5-NS5* Probe}\label{sec:KK5NS5probe}
%%------------------------------------------------

The actions for Kaluza-Klein branes in string/M-theory have been discussed in Refs.~\cite{Bergshoeff:1997gy}-\cite{Eyras:1998hn}. For the heterotic string, the KK5 probe action is given by \cite{Bergshoeff:1997gy,Janssen:1999pa}
\be
S_{\rm KK5} = - \mu_5 \int d^6\sigma e^{-2\phi} k^2 
\sqrt{-\det (\Pi_{ab}+k^{-2}F_a F_b)}
+ S_{\rm WZ}.
\label{eq:KK5}
\ee
Here, $g_{ab}$ is the induced metric; the pull-back of tensors is always denoted with indices $a,b,\ldots$. Also,
\bea
k^2 &=& k^{\mu} k^{\nu} G_{\mu\nu} = G_{zz}, \nonumber \\
\Pi_{ab} &=& g_{ab} - k^{-2} k_a k_b, \nonumber \\
F_a &=& \del_a s + \del_a X^\mu k^\nu B_{\mu\nu},
\eea
where $k \equiv (\del_z)^\mu$ is the Killing vector associated to the isometry direction $z$ of the probe KK-brane, and the field $s$ is a world-volume scalar.
The factor $k^2$ in the kinetic term is a characteristic of the Kaluza-Klein monopole and accounts for the factor $R^2$ of the tension in Eq.~\eq{NS5KK5tension}. 
The actions are proportional to $\mu_5$; the string coupling $g_s$ is included in the background solutions \eq{NS1W} and \eq{KK5NS5}. Some references often include $g_s$ in tensions, not in  backgrounds ({\it e.g.}, Ref.~\cite{Big}).
When the background metric has an off-diagonal element between the isometry direction $z$ 
and a direction parallel to the probe brane, $k_a \neq 0$ in general. This is actually the case for the W*-background for the configuration \eq{heterotic}. 

The relevant term of the Wess-Zumino action is 
\be
S_{\rm WZ} = \mu_5 \int d^{6}\sigma \, \frac{\sqrt{-g}}{6!} 
\epsilon^{a_0 \cdots a_5} \, k^\mu 
\del_{a_0}X^{\mu_0} \cdots \del_{a_5}X^{\mu_5} \, 
A_{\mu \mu_0 \cdots \mu_5}
\ee
(the full Wess-Zumino action is given in Ref.~\cite{Janssen:1999pa}), where $dA_7 = k^2 e^{-2\phi} *dA_1$ or 
\bea
F_{\mu_1 \cdots \mu_8} &=& 
\frac{1}{2!} \,  k^2 e^{-2\phi} \epsilon_{\mu_1 \cdots \mu_{10}} 
F^{\mu_9 \mu_{10}}, \\
A_{\mu} &=& k^{-2} k^{\nu} G_{\mu\nu}.
\eea

The NS5 probe action is given by \cite{Callan:1991ky}
\be
S_{\rm NS5} = - \mu_5 \int d^6\sigma e^{-2\phi} \sqrt{-\det g_{ab}}
+ \mu_5 \int B_6,
\label{eq:NS5}
\ee
where $dB_6 = e^{-2\phi}*dB_2$. In principle, the effective NS5*-brane action could be derived from higher order corrections of the Kaluza-Klein monopole action in a manner similar to the effective D6-D2* action. Here, we take the actions \eq{NS5} and \eq{KK5} as our starting point since these corrections are less known.

For the KK5-NS5* background, one can choose a static gauge:
\bea
X^0 &=& \tau, \nonumber \\
X^9 &=& 0, \nonumber \\
X^{a+3} &=& \sigma^a, \qquad a=1,\ldots,5, \nonumber \\
%X: 0->2\pi R
%R \sigma
X^i &=& X^i(\tau).
\eea
Using the small velocity approximation, we get the KK5-NS5* probe action as
\be
{\cal L} = - v_3 (\tau_{\rm KK5} + \tau_{\rm NS5^*}) 
+ \frac{1}{2} \, v_3 (\tau_{\rm KK5} f_5 + \tau_{\rm NS5^*} f_K) v^2 + \cdots.
\label{eq:KK5NS5inKK5NS5}
\ee

For the NS1-W* background, the gauge choice and ansatz are
\bea
X^0 &=& \tau, \nonumber \\
X^9 &=& \tilde{y}(\tau), \nonumber \\
X^{a+3} &=& \sigma^a, \qquad a=1,\ldots,5, \nonumber \\
%X: 0->2\pi R
%R \sigma
X^i &=& X^i(\tau).
\eea
The $\tilde{y}$ is again a cyclic coordinate, so there is a conserved momentum $\tilde{P}$. Since we want the probe KK5-NS5* brane to carry no momentum, $\tilde{P}=0$. Substituting this back to the action, we get the KK5-NS5* probe action as
\be
{\cal L} = 
- v_3 \frac{\tau_{\rm KK5} f_n + \tau_{\rm NS5^*} f_1}{\sqrt{f_1 f_n}} 
+ \frac{1}{2} \, v_3 (\tau_{\rm KK5} f_n + \tau_{\rm NS5^*} f_1) \sqrt{f_1 f_n} v^2 + \cdots.
\label{eq:KK5NS5inNS1W}
\ee
Again the results \eq{KK5NS5inKK5NS5} and \eq{KK5NS5inNS1W} agree with the heuristic analyses. 

As a check, the probe results we obtained \eq{NS1WinNS1W}, \eq{NS1WinKK5NS5}, \eq{KK5NS5inKK5NS5}, and \eq{KK5NS5inNS1W} are consistent with various dualities. For instance, consider the D6-D2* background in the type IIA string. The effective particle action from the D4-D0* probe is given by
\be
{\cal L}_{\rm D4-D0^*} =-\frac{\mu_4 \VA Z_2 - \mu_0 Z_6}{\gA \sqrt{Z_2 Z_6}} 
+ \frac{1}{2 \gA} (\mu_4 \VA Z_2 - \mu_0 Z_6) \sqrt{Z_2 Z_6} v^2 + \cdots.
\label{eq:D4D0inD6D2}
\ee
The effective membrane action from the D6-D2* probe is given by
\be
{\cal L}_{\rm D6-D2^*} = -\frac{\mu_6 \VA - \mu_2}{\gA} 
+ \frac{1}{2 \gA} (\mu_6 \VA Z_2 - \mu_2 Z_6) v^2 + \cdots.
\label{eq:D6D2inD6D2}
\ee
Equations~\eq{NS1WinKK5NS5} and \eq{KK5NS5inKK5NS5} are S-dual to \eq{D4D0inD6D2} and \eq{D6D2inD6D2}, respectively. The heterotic results are also invariant under the T-duality $R \rightarrow 1/R$.

%%%%%%%%%
\section*{Acknowledgments}
%%%%%%%%%

We would like to thank Gungwon Kang, Kengo Maeda, Don Marolf, Joe Polchinski, Masa-aki Sakagami, Makoto Sakaguchi, and Jiro Soda for useful discussions. I would also like to thank the participants of ``Working group on singularity in string and general relativity" for useful discussions. This work was supported in part by a Grant-in-Aid for Scientific Research (13740167) from the Ministry of Education, Culture, Sports, Science and Technology, Japan.

%\pagebreak

\appendix

%%%%%%%%%
\section{The p.p. Singularity in the NS1-W* Geometry}\label{sec:appA}
%%%%%%%%%

The p.p. singularity has been discussed in various works; see, {\it e.g.}, Refs.~\cite{Kaloper:1997hr,Horowitz:1997si} for recent discussions in the context of string theory. Standard textbooks in general relativity often have a treatment; see, {\it e.g.}, Chap.~31.2 of Ref.~\cite{MTW} for an elementary introduction. We mostly follow the notations and the derivation of Ref.~\cite{Kaloper:1997hr}. Many equations are unchanged, but the final result is not the same. This is because the reference focuses on the standard branes whose geometry could be singular only at $r=0$.

A p.p. singularity is the one where the Riemann tensor components are unbounded in a p.p. frame along at least one non-spacelike curve. The calculation consists of three steps: (i) First, choose a timelike geodesic (an infalling observer) along which we calculate the Riemann tensor; (ii) Then, choose a convenient frame to calculate the Riemann tensor, and find the transformation which relates the frame to the observer's frame; (iii) Finally, transform the Riemann tensor to the observer's frame to obtain the observer's tidal forces.

{\it (i) Geodesics:}
We denote the metric as 
\be
ds^2 = 2F_2 \, du dv + F_3^2 du^2 + F_1^{-2} dx^idx^i,
\ee
where $z=u+v$ and $t=v$. For the NS1-W* geometry, $F_1=1, F_2=f_1^{-1}$, and $F_3^2=f_1^{-1}f_n$. The geodesic equation leads to the ten-velocity of the observer $V^{\mu}=dx^{\mu}/d\tau$, where
\bea
\frac{du}{d\tau} &=& \frac{P-E}{F_2}, \qquad
\frac{dv}{d\tau} = \frac{P}{F_2}+\left(\frac{F_3}{F_2}\right)^2(E-P)^2, \\
(\frac{dr}{d\tau})^2 &=& 
(E-P)^2 \left(\frac{F_1F_3}{F_2}\right)^2+\frac{2F_1^2}{F_2}(E-P)P-F_1^2.
\eea
Here, $E$ and $P$ are the integrals of motion. We also fix the motion in the $u-r$ plane only, which is consistent with the other components of the geodesic equation.

{\it (ii) Lorentz transformation:}
The Riemann tensor $R^{klmn}$ may be first computed not in the observer's frame [with coordinates $(\mt,\mz,\mr,\ldots)$], but in a stationary orthonormal frame [with coordinates $(\lt,\lz,\lr,\ldots)$]:
\be
e^{\lt} = \frac{F_2}{F_3} dv, \qquad
e^{\lz} = F_3 du + \frac{F_2}{F_3} dv, \qquad
e^{\lr} = F_1^{-1} dr.
\ee
The Riemann tensor in the observer's frame is then obtained by a Lorentz boost. For the NS1-W* geometry, the Riemann tensor components diverge as $r \rightarrow r_n$ at this level. The components may not diverge for other p.p. singularities, and the singularity may appear as the result of the boost. 

The ten-velocity in this frame is given by $V^{a}=e^a_{~\mu}V^{\mu}$, where 
\bea
V^{\lt} &=& \frac{F_3}{F_2}(E-P) + \frac{P}{F_3}, \qquad
V^{\lz} = \frac{P}{F_3}, \\
V^{\lr} &=& \pm \left\{
\left(\frac{F_3}{F_2}\right)^2(E-P)^2 + \frac{2(E-P)P}{F_2}-1 
\right\}.
\eea
Now, the Lorentz transformation we need takes a unit timelike vector $N^a = \delta^a_{~\lt}$ (the observer's ten-velocity in his own frame) into $V^a$ (the ten-velocity in the stationary frame), so $V^{a}=L^a_{~b} N^b$. One simple choice of such a SO(1,9) matrix is 
\be
L^a_{~b} = \left(
\begin{array}{cccc}
V^{\lt} & V^{\lz} & V^{\lr} & {\lt} \\
V^{\lz} & 1+\frac{(V^{\lz})^2}{V^{\lt}+1} 
		& \frac{V^{\lz} V^{\lr}}{V^{\lt}+1} & 0 \\
V^{\lr} & \frac{V^{\lz} V^{\lr}}{V^{\lt}+1} 
		& 1+\frac{(V^{\lr})^2}{V^{\lt}+1} & 0 \\
0 & 0 & 0 & 1 
\end{array}
\right),
\label{eq:Lorentz}
\ee
where 1 is an identity matrix. Strictly speaking, the frame obtained by the transformation may differ from the parallelly propagated frame by a rotation, but it suffices to see the singularity.

{\it (ii) Tidal forces:}
Finally, apply the Lorentz transformation \eq{Lorentz} to the Riemann tensor in the stationary frame $R^{klmn}$; 
$R^{abcd} = L^a_{~k} L^b_{~l} L^c_{~m} L^d_{~n} R^{klmn}$.
This gives the tidal forces for the observer.
For the NS1-W* background, the leading terms as $r \rightarrow r_n$ for some components are
\be
R^{\mz\mr\mz\mr}, \hat{R}^{\mt\mr\mt\mr}, \hat{R}^{\mz\mr\mt\mr} 
\rightarrow \frac{2P^2(r_1^3+4r_n^3)}{3r_n^3(r-r_n)^2}, 
\ee
\be
R^{\mt\mz\mt\mz} \rightarrow 
\frac{-2(r_1^3+4r_n^3)\{r_n^3+2P^2(r_1^3+r_n^3)-2PE(r_1^3+r_n^3) \}}{r_n^4(r_1^3+r_n^3)(r-r_n)}.
\ee
This shows that the NS1-W* geometry has a p.p. curvature singularity at $r=r_n$.

%%%%%%%%%
\section{Repulson and Naked Singularity}\label{sec:appB}
%%%%%%%%%

Here, we limit ourselves to a general static spherically symmetric metric in $n$-dimensions:
\be
ds^2 = -f^2(r) dt^2 + f(r)^{-2} dr^2 + R(r)^2 d\Omega_{n-2}^2.
\ee
Then, the repulsive singularities must be timelike. The argument goes as follows. From Eq.~\eq{potential}, the effective potential is $V_{\rm eff} =f^2$. So, an infinite repulsive potential means that $f^2 \rightarrow \infty$ or $G_{00} \rightarrow -\infty$ as $r \rightarrow 0$, where we chose that $r=0$ is the singularity. Thus, $f$ never has a zero near the singularity. Now, define a new radial coordinate $r_*=\int f^{-2} dr$ and null coordinates $u=t-r_*$, $v=t+r_*$. Then the two-dimensional part of the metric reads as $ds_2^2 = -f^2 dudv$. Now, the causal structure of the singularity is as follows:
\begin{enumerate}
\item If one can choose the integration constant so that $r_*=0$ at $r=0$ and if $f^2>0$, then the singularity is timelike.
\item Same as above, but if $f^2<0$, it is spacelike.
\item If $r_*=-\infty$ at $r=0$, it is null.
\end{enumerate}
Because $f$ never has a zero near the singularity and $f^2>0$ in our case, it is timelike. 
Therefore, it must be either locally naked, {\it i.e.}, surrounded by a horizon, or globally naked, {\it i.e.}, a true naked singularity.

Even though repulsive singularities are timelike, the converse is not true. The standard D6-brane has a timelike singularity which is attractive. This is because repulsive behavior is not necessary in order to have $r_*=0$ at the singularity. One could have $f \sim r^a$, where $a<1/2$ (for the D6-brane, $a=1/4$). Another example is the cosmic string with a conical space.

\begin{figure}
\begin{center}
\hspace{.25in}\epsfbox{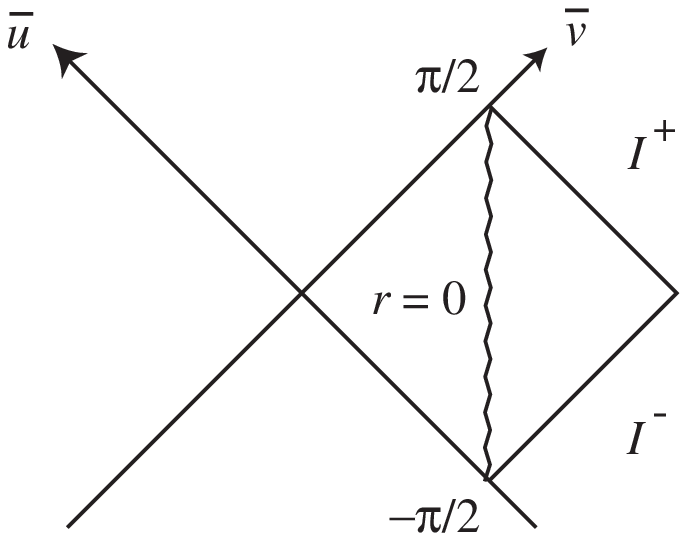}
\caption{The Penrose diagram in the local patch.}
\label{fig:timelike}
\end{center}
\end{figure}

The above causal structures can be checked by drawing the resulting Penrose diagram. The new null coordinates 
\be
\tan \bar{u} = -e^{-u}, \quad \tan \bar{v} = e^{v}, \quad
\mbox{where } |\bar{u}|<\pi/2, |\bar{v}|<\pi/2
\ee
transforms the metric into 
\be
ds_2^2 = -f^2 \sec^2 \bar{u} \sec^2 \bar{v} d\bar{u}d\bar{v}.
\ee 
When $f^2>0$, the Penrose diagram then looks like \fig{timelike} in the local patch.

\end{document}